\documentclass[prc,aps,twocolumn,floats,floatfix,showpacs,superscriptaddress,nofootinbib]{revtex4}
\usepackage{amsmath}
\usepackage{amsfonts}
\usepackage{graphicx}

\begin{document}

\title{Proton-neutron symmetry in $^{92}$Zr, $^{94}$Mo with Skyrme interactions in a separable approximation}

\author{A. P. Severyukhin}
\affiliation{Bogoliubov Laboratory of Theoretical Physics,
             Joint Institute for Nuclear Research,
             141980 Dubna, Moscow region, Russia}
\author{N. N. Arsenyev}
\affiliation{Bogoliubov Laboratory of Theoretical Physics,
             Joint Institute for Nuclear Research,
             141980 Dubna, Moscow region, Russia}
\author{N. Pietralla}
\affiliation{Institut f\"{u}r Kernphysik,
             Technische Universit\"{a}t Darmstadt,
             64289 Darmstadt, Germany}

\begin{abstract}
Starting from a Skyrme interaction we study the properties of the
low-energy spectrum of quadrupole excitations in $^{90,92}$Zr,
$^{92,94}$Mo. The coupling between one- and two-phonon terms in
the wave functions of excited states are taken into account. We
use the finite-rank separable approximation which enables one to
perform the QRPA calculations in very large two-quasiparticle
spaces. Our results from the SGII Skyrme interaction in connection
with the volume pairing interaction are in reasonable agreement
with experimental data. In particular, we present the successful
description of the $M1$ transition between low-energy quadrupole
excitations.
\end{abstract}


\date{01.06.2012}

\maketitle
%
%
\section{Introduction}
Properties of the quadrupole-collective isovector excitations of
the valence shell of heavy nuclei reflect three main aspects:
collectivity, shell structure, and the isospin degree of freedom.
The balance between these aspects has been studied within a series
of experiments on so-called mixed-symmetry (MS) states. There is a
rather complete list of references on that subject in
review~\cite{pbl08}. These isovector excitations were predicted in
the proton-neutron version of the interacting boson model
(IBM-2)~\cite{ibmbook}, where the proton-neutron symmetry of the
wave functions is quantified by the bosonic analog of isospin, F
spin~\cite{aoit77,oai78,iach84}. In particular, there are the
symmetric states with maximum F-spin ($F=F_{max}$) and the MS ones
with $F<F_{max}$. One of the well-known examples is the MS
quadrupole state with $F=F_{max}-1$ in the nucleus $^{94}$Mo which
was indentified by the measured strong $M1$ transitions and
weakly-collective $E2$ transitions to low-lying symmetric
states~\cite{P99,F03,B07}. In contrast to the case for $^{94}$Mo,
the next lighter even-even $N=52$ isotone, $^{92}$Zr has the
proton subshell closure $Z=40$ and the collectivity of the $2^+_1$
state decreases. As a result, the collective MS structure and the
structures showing the considerable F-spin breaking have been
observed experimentally~\cite{W02,F05,W11}. This sensitivity makes
the MS states highly appealing objects for studies in microscopic
approaches.

The results of the large-scale shell-model
calculations~\cite{W02,F05,L00,H07} have indicated the occurrence
of MS quadrupole states in $^{94}$Mo, $^{92}$Zr and the breaking
of F-spin symmetry in $^{92}$Zr. For the case of $^{94}$Mo, the
wave functions, the $M1$ and $E2$ matrix elements support the MS
assignments for the $2^+_3$ state and the consistency with the
IBM-2 predictions proves the collective character of the observed
low-lying levels. For $^{92}$Zr, the calculated $2^+_{1,2}$ states
are predominantly the pure neutron and isovector excitations,
respectively. This means that the single particle and collective
degrees of freedom are present in the low-energy spectrum.

An alternative powerful microscopic approach is the
quasiparticle-phonon model (QPM)~\cite{solo}. The model
Hamiltonian is diagonalized in a space spanned by states composed
of one, two and three phonons which are generated in quasiparticle
random phase approximation (QRPA). The separable form of the
residual interaction is the practical advantage of the QPM which
allows one to perform the structure calculations in large
configurational spaces. For $^{94}$Mo, the QPM confirms the IBM-2
level scheme, selection rules and, in particular, the dominant
one-phonon structure of the transitions to the first and third
$2^+$ states~\cite{B07,LS00}. Also, the QPM gives a satisfactory
and comprehensive description of the large variety of data
measured for $^{92}$Zr~\cite{LS04,LS06,W11}. The QPM wave
functions of the two lowest $2^+$ states are dominated by the
lowest neutron and proton two-quasiparticle components and the
structures are very similar to results of the shell model
calculations. Thus, the QPM can provide a microscopic support to
the IBM-2 scheme. However, it is difficult to extrapolate the
model parameters to new regions of nuclei.

The properties of MS states are particularly sensitive to the
proton-neutron interaction~\cite{A09,C10}. On the other hand, the
dominance of the proton-neutron attraction is one of the main
characteristics of the effective nucleon-nucleon interaction. It
can be traced back to the cooperation of its $T=0$ and $T=1$
channels. This is a good possibility to examine microscopic
approaches using effective nucleon-nucleon interactions. One of
the successful tools for nuclear structure studies is the QRPA
with the Skyrme interaction~\cite{ter05,rev2007}. These QRPA
calculations allow to relate the properties of the ground states
and excited states through the same energy density functional.
Although such an approach describes the properties of the
low-lying states less accurately than more phenomenological ones,
the results are still in a reasonable agreement with experimental
data, in particular, with respect to qualitative features of the
properties of the $2^+_1$ states of predominantly one-phonon
excitations, see for example~\cite{KSGG02,colo03,ter06,ter11}.

The complexity of calculations taking into account the coupling
between one-phonon and more complex states increases rapidly with
the size of the configurational space. Making use of the finite
rank separable approximation (FRSA) \cite{gsv98,svg08} for the
residual interaction enables one to perform QRPA calculations in
very large two-quasiparticle spaces. Taking into account the basic
QPM ideas, the approach has been generalized to take into account
the coupling between one- and two-phonon components of the wave
functions~\cite{svg04}. The FRSA has been used to study the
electric low-lying states and giant resonances within the QRPA and
beyond~\cite{gsv98,svg08,svg04,ssvg02,savg09}. Alternative schemes
to factorize the residual interaction have also been considered in
Refs.~\cite{SS81,sar99,nest02}.

Before to investigate the occurrence of MS states in neutron-rich
nuclei one needs to be sure that the approach is sufficiently good
to reproduce characteristics of the low-energy spectrum of
quadrupole excitations of nuclei in the mass range $A\approx90$,
in particular, the $M1$ transitions between the excitations. This
initial paper gives an illustration of our approach for $^{92}$Zr,
$^{94}$Mo in comparison to the $N=50$ isotones $^{90}$Zr,
$^{92}$Mo with closed neutron shell. Preliminary results of our
studies for $^{94}$Mo are reported already in Ref.~\cite{savpg11}.

This paper is organized as follows. In Sec.~II, we sketch the
method allowing to consider the phonon-phonon coupling. In
particular, the QRPA equations in the case of the finite rank
separable form of the residual interaction are discussed in
Sec.~IIA. The coupling between one- and two-phonon terms in the
wave functions of excited states are taken into account in
Sec.~IIB. In Sec. III, we discuss the details of our calculations
and show how this approach can be applied to treat the
proton-neutron mixed-symmetry states. Results of our calculations
for properties of the quadrupole states in $^{90,92}$Zr,
$^{92,94}$Mo are given in Sec.IV. Sec.~IVA is devoted to QRPA
analysis, while the effects of the phonon-phonon coupling are
discussed in Sec.~IVB. Conclusions are finally drawn in Sec.~V.
%
%
\section{The method}
This method has already been introduced in
Refs.~\cite{gsv98,ssvg02,svg08,svg04}; hence, let us briefly
summarize the different steps. For the present study, the
anharmonicity of the low-energy vibrations is  constrained by the
coupling between one- and two-phonon terms in the wave functions
of excited states.

\subsection{Implementation of QRPA}
The starting point of the method  is the HF-BCS
calculation~\cite{RingSchuck} of the ground state, where spherical
symmetry is imposed on the quasiparticle wave functions. The
continuous part of the single-particle spectrum is discretized by
diagonalizing the Skyrme HF Hamiltonian on a harmonic oscillator
basis. As effective interactions, Skyrme interaction are used in
the particle-hole (p-h) channel, and the pairing correlations are
generated by a density-dependent zero-range force
\begin{equation}
V_{pair}({\bf r}_1,{\bf r}_2)=V_{0}\left( 1-\eta\left(\frac{\rho
\left( r_{1}\right) } {\rho _{0}}\right)^{\alpha}\right) \delta
\left( {\bf r}_{1}-{\bf r}_{2}\right), \label{pair}
\end{equation}
where $\rho \left( r_{1}\right)$ is the particle density in
coordinate space; $\rho _{0}$ is equal to the nuclear saturation
density;  $\alpha$, $\eta$ and $V_{0}$ are model parameters. We
use $\alpha$=1; $\eta$= 0, $\eta$= 0.5, and $\eta$= 1 are the
cases of a volume interaction, a mixed interaction and a
surface-peaked interaction, respectively. The strength $V_{0}$ is
fitted to reproduce the odd-even mass difference in the region of
nuclei considered here. In order to limit the pairing
single-particle space, we have used the smooth cutoff at 10 MeV
above the Fermi energies~\cite{svg08,B85,K90}.

The residual interaction in the p-h channel $V^{ph}_{res}$ and in
the p-p channel $V^{pp}_{res}$ can be obtained as the second
derivative of the energy density functional with respect to the
particle density and the pair density, accordingly. As proposed in
Ref.~\cite{gsv98}, we simplify $V^{ph}_{res}$ by approximating it
by its Landau-Migdal form. All Landau parameters with $l > 1$ are
equal to zero in the case of Skyrme interactions. We keep only the
$l=0$ terms in $V^{ph}_{res}$. In this work we study only normal
parity states and one can neglect the spin-spin terms since they
play a minor role~\cite{KSGG02,ssvg02}. Also, the two-body Coulomb
and spin-orbit residual interactions are dropped. Therefore we can
write the residual interaction as
\begin{equation}
\label{res.int} V^{a}_{res}({\bf r}_1,{\bf r}_2) = N_0^{-1}[
F_0^{a}(r_1)+ F_0^{'a}(r_1){\bf \tau }_1\cdot{\bf \tau }_2] \delta
({\bf r}_1 - {\bf r}_2)
\end{equation}
where $a$ is the channel index $a=\{ph,pp\}$; ${\bf \tau}_i$ are
the isospin operators, and $N_0 = 2k_Fm^{*}/\pi^2\hbar^2$ with
$k_F$ and $m^{*}$ standing for the Fermi momentum and nucleon
effective mass.

The expressions for $F^{ph}_0, F^{'ph}_0$ and $F^{pp}_0,
F^{'pp}_0$ can be found in Ref.\cite{sg81} and in
Ref.~\cite{svg08}, respectively. Since the definition of the
pairing force~(\ref{pair}) involves the energy cutoff of the
single-particle space to restrict the active pairing space within
the mean-field approximation, this cutoff is still required to
eliminate the p-p matrix elements of the residual interaction in
the case of the subshells that are far from the Fermi energies as
in Ref.~\cite{svg08}.

The p-h matrix elements and the antisymmetrized p-p matrix
elements can be written as the separable form in the angular
coordinates~\cite{gsv98,svg08}. After integrating over the angular
variables we use a $N$-point integration Gauss formula for the
radial integrals. Thus, the residual interaction can be expressed
as the sum of $N$ terms in FRSA for the Skyrme residual
interaction~\cite{gsv98,svg08}.

We introduce the phonon creation operators
\begin{eqnarray}
\label{phonon} Q_{\lambda \mu i}^{+}= \frac 12\sum_{jj'}\left( X
_{jj'}^{\lambda i}\,A^{+}(jj';\lambda \mu)\right.
\nonumber\\
\left. -(-1)^{\lambda -\mu }Y _{jj'}^{\lambda i}\,A(jj';\lambda
-\mu )\right),
\end{eqnarray}
\begin{equation}
A^{+}(jj';\lambda \mu )=\sum_{mm'} \langle jmj'm' \mid \lambda \mu
\rangle \alpha _{jm}^{+}\alpha _{j'm'}^{+},
\end{equation}
where $\lambda $ denotes the total angular momentum and $\mu $ is
its z-projection in the laboratory system. $\alpha^{+}_{jm}$
($\alpha_{jm}$) is the quasiparticle creation (annihilation)
operator and $jm$ denote the quantum numbers $nljm$. One assumes
that the ground state is the QRPA phonon vacuum $\mid 0\rangle $
and the one-phonon excited states are $Q_{\lambda \mu
i}^{+}\mid0\rangle$ with the normalization condition
\begin{eqnarray}
\langle 0\mid Q_{\lambda \mu i} Q_{\lambda' \mu' i'}^{+} \mid
0\rangle
\nonumber\\
=\delta _{\lambda\lambda'} \delta_{\mu\mu'}\frac
12\sum_{jj^{'}}\left( X _{jj'}^{\lambda i} X _{jj'}^{\lambda i'} -
Y _{jj'}^{\lambda i} Y _{jj'}^{\lambda i'}\right)=  \delta _{ii'}.
\label{phononnorm}
\end{eqnarray}
Making use of the linearized equation-of-motion approach one can
get the QRPA equations~\cite{RingSchuck}. Solutions of this set of
linear equations yield the one-phonon energies $\omega _{\lambda
i}$ and the amplitudes $X_{jj'}^{\lambda i}$, $Y_{jj'}^{\lambda
i}$ of the excited states. There are two types of QRPA matrix
elements: the $A^{(\lambda)}_{(j_1j'_1)(j_2j'_2)}$ matrix related
to forward-going graphs and the
$B^{(\lambda)}_{(j_1j'_1)(j_2j'_2)}$ matrix related to
backward-going graphs, and the dimension of these matrices is the
space size of the two-quasiparticle configurations. Using the FRSA
for the residual interaction, the eigenvalues of the QRPA
equations can be obtained as the roots of a relatively simple
secular equation, where the matrix dimensions never exceed $6N
\times 6N$ independently of the two-quasiparticle configuration
space size~\cite{gsv98,svg08}. The studies
\cite{ssvg02,svg08,svg04} enable us to conclude that $N$=45 for
the rank of our separable approximation is enough for the electric
excitations ($J \le 6 $) in nuclei with $A\le 208$.
%
%
\subsection{Phonon-phonon coupling}
In the next stage, we construct the wave functions from a linear
combination of one-phonon and two-phonon configurations
\begin{eqnarray}
\Psi _\nu (\lambda \mu ) = \left(\sum_iR_i(\lambda \nu )Q_{\lambda
\mu i}^{+}\right.
\nonumber\\
\left.+\sum_{\lambda _1i_1\lambda _2i_2}P_{\lambda _2i_2}^{\lambda
_1i_1}(\lambda \nu )\left[ Q_{\lambda _1\mu _1i_1}^{+}Q_{\lambda
_2\mu _2i_2}^{+}\right] _{\lambda \mu }\right)|0\rangle.
\label{wf2ph}
\end{eqnarray}
with the normalization condition
\begin{equation}
\sum\limits_iR_i^2(\lambda \nu)+ 2\sum_{\lambda _1i_1 \lambda
_2i_2} (P_{\lambda _2i_2}^{\lambda _1i_1}(\lambda \nu))^2=1.
\label{norma2ph}
\end{equation}
Because of the completeness and orthogonality conditions for the
phonon operators, the bifermion operators $A^{+}(jj';\lambda \mu
)$ and $A(jj';\lambda \mu)$ can be expressed by the phonon ones.
The Hamiltonian can be rewritten in terms of quasiparticle and
phonon operators~\cite{solo,svg04}.

The amplitudes $R_i(\lambda \nu)$ and
$P_{\lambda_2i_2}^{\lambda_1i_1}(\lambda \nu)$ are determined from
the variational principle
\begin{eqnarray}
\delta \left( \langle \Psi_\nu (\lambda \mu )\mid H \mid \Psi_\nu
(\lambda \mu )\rangle \right.
\nonumber\\
\left. -E_\nu \left( \langle {\Psi _\nu (\lambda \mu )|\Psi _\nu
(\lambda \mu )\rangle - 1}\right) \right) = 0,
\end{eqnarray}
which leads to a set of linear equations~\cite{solo,svg04}
\begin{eqnarray}
(\omega _{\lambda i}-E_\nu )R_i(\lambda \nu )
\nonumber\\
+\sum_{\lambda _1i_1 \lambda_2i_2} U_{\lambda _2i_2}^{\lambda
_1i_1}(\lambda i) P_{\lambda_2i_2}^{\lambda _1i_1}(\lambda \nu
)=0,
\label{2pheq1}
\end{eqnarray}
\begin{eqnarray}
2(\omega _{\lambda _1i_1}+\omega _{\lambda _2i_2}-E_\nu
)P_{\lambda _2i_2}^{\lambda _1i_1}(\lambda \nu)
\nonumber\\
+\sum\limits_i U_{\lambda _2i_2}^{\lambda _1i_1}( \lambda
i)R_i(\lambda \nu )=0.
\label{2pheq2}
\end{eqnarray}
The Pauli principle corrections are dropped here since the effect
on the lowest excited states is small~\cite{solo}.

The rank of the set of linear equations (\ref{2pheq1}),
(\ref{2pheq2}) is equal to the number of one- and two-phonon
configurations included in the wave function (\ref{wf2ph}). Its
solution requires to compute the matrix elements coupling one- and
two-phonon configurations
\begin{equation}
U_{\lambda _2i_2}^{\lambda _1i_1}(\lambda i)= \langle 0|
Q_{\lambda i } H \left[ Q_{\lambda _1i_1}^{+}Q_{\lambda
_2i_2}^{+}\right] _{\lambda} |0 \rangle.
\end{equation}
Eqs.~(\ref{2pheq1}), (\ref{2pheq2}) have the same form as the QPM
equations~\cite{solo}, but the single-particle spectrum and the
parameters of the residual interaction are calculated with the
Skyrme forces.
\begin{figure*}[t!]
\includegraphics[width=2.0\columnwidth]{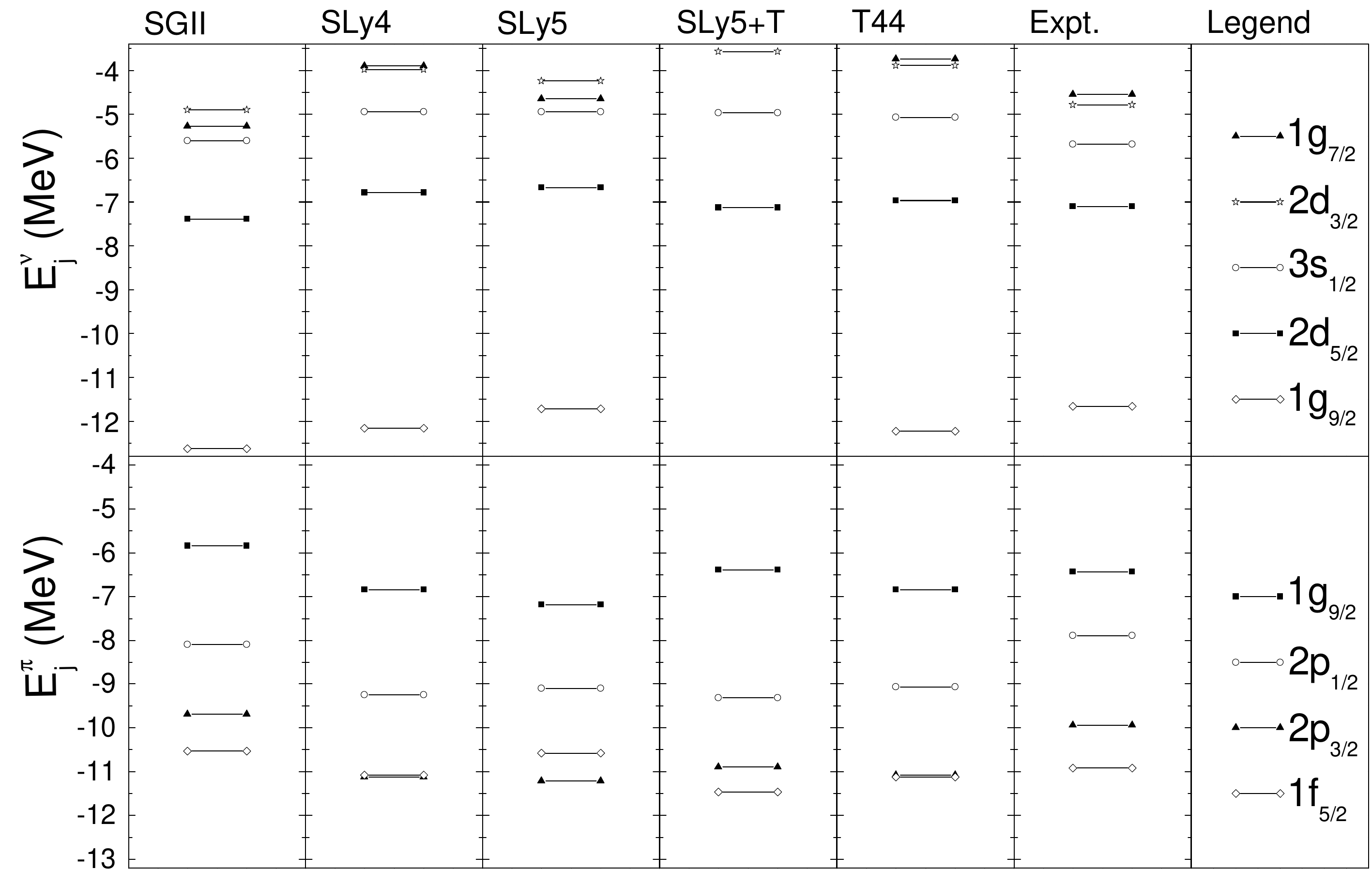}
\caption{Neutron (top) and proton (bottom) single-particle
energies (in MeV) near the Fermi energies for $^{92}$Zr calculated
with SGII, SLy4, SLy5, SLy5+T, and T44. The experimental spectra
are taken from Ref.~\cite{M79}.}
\end{figure*}
\begin{figure*}[t!]
\includegraphics[width=2.0\columnwidth]{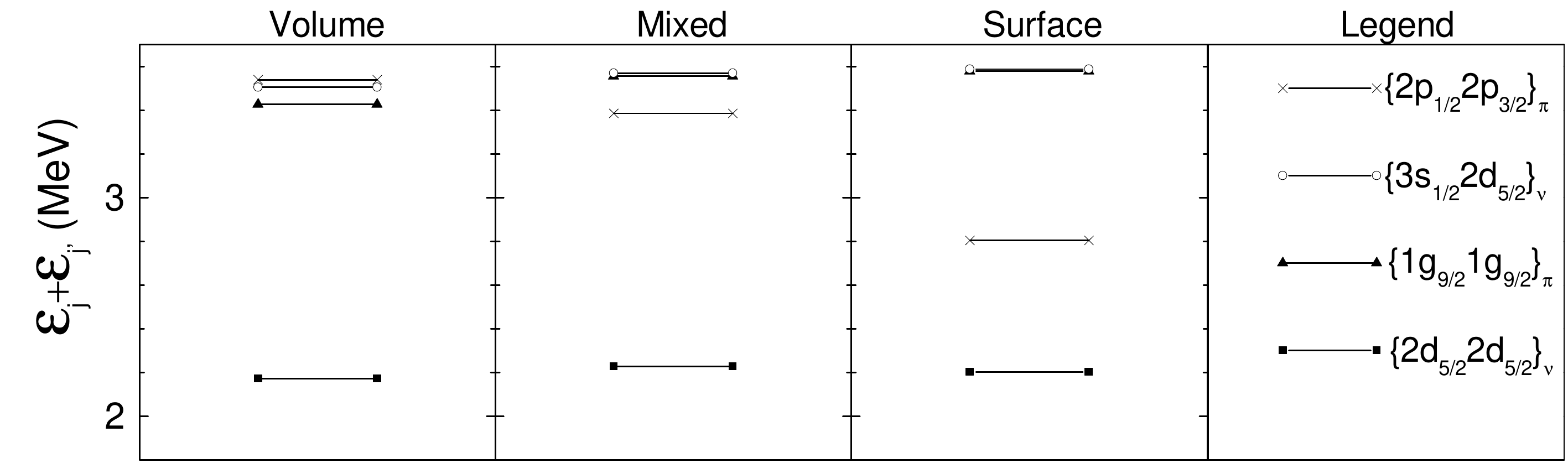}
\caption{Energies (in MeV) of lowest two-quasiparticle states in
$^{92}$Zr. The results of calculations with the volume, the mixed,
and the surface-peaked pairing interactions are shown.}
\end{figure*}
\begin{figure}[t!]
\includegraphics[width=1.0\columnwidth]{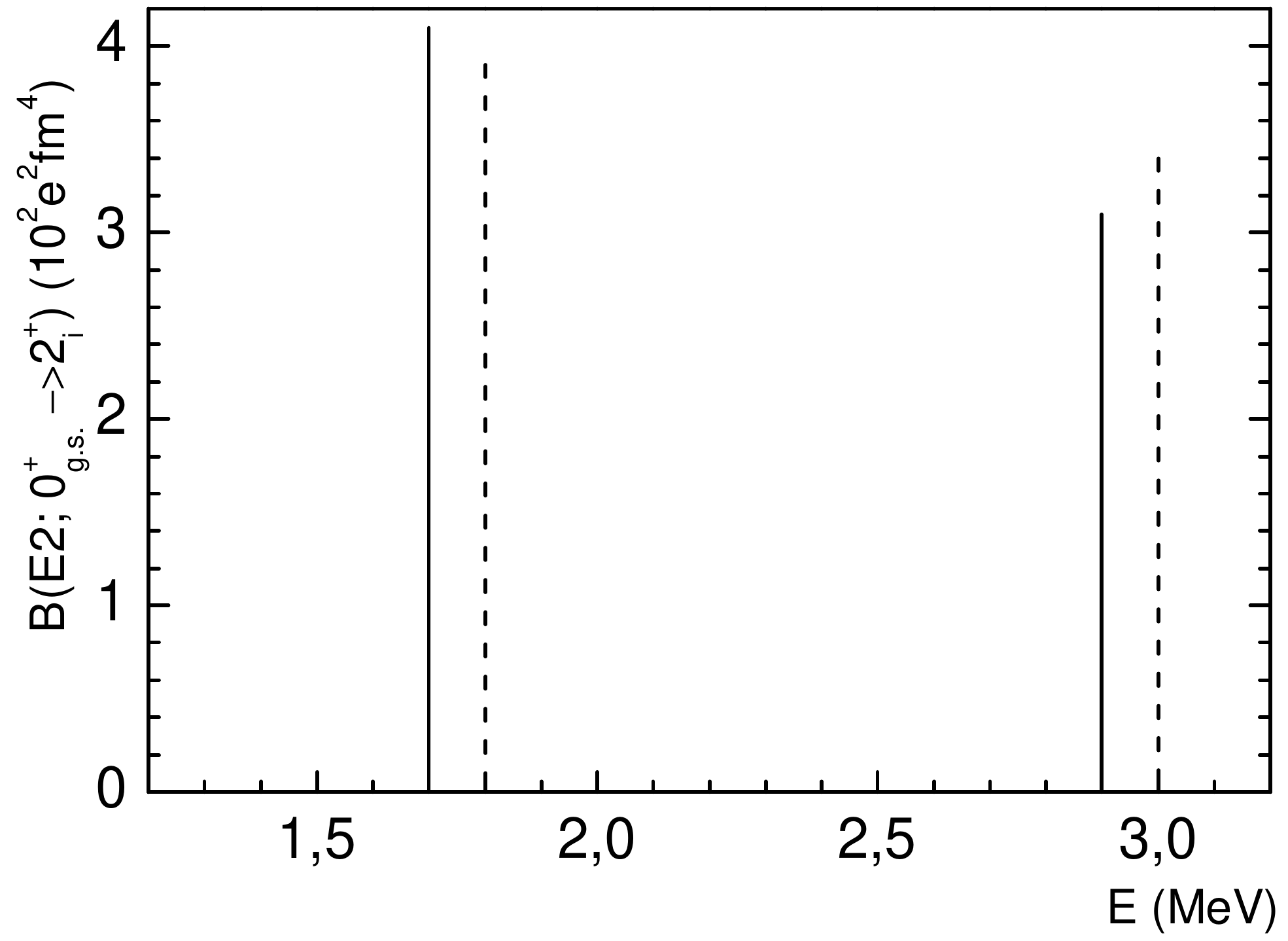}
\caption{Energies and B(E2)-values of the lowest quadrupole states
in $^{92}$Zr. The QRPA calculations are performed without (dashed
line) and with (solid line) the residual p-p interaction.}
\end{figure}
%
%
%
\section{Details of calculations}
We apply the approach to study the low-energy spectrum of $2^+$
excitations in the nuclei $^{90,92}$Zr, $^{92,94}$Mo. In our
calculations the single-particle continuum is discretized by
diagonalizing the HF Hamiltonian on a basis of 12 harmonic
oscillator shells and cutting off the single-particle spectra at
the energy of 100~MeV. This is sufficient to exhaust practically
all the energy-weighted sum rule. Because of the large
configurational space, we do not use effective charges.

The single-particle structure around the Fermi level is a key
ingredient in the microscopic analysis. For the investigation, we
adopt five Skyrme interactions, namely, SGII~\cite{sg81},
SLy4~\cite{sly}, SLy5~\cite{sly}, SLy5+T~\cite{csfb07}, and
T44~\cite{T44}. The SGII parameterization is the successful
attempt to describe the spin-dependent properties from a standard
Skyrme force. In particular, one obtains a good description of
experimental energies of the Gamow-Teller resonance of
$^{90}$Zr~\cite{sg81}. The parameters of the forces SLy4 and SLy5
have been adjusted to reproduce nuclear matter properties, as well
as nuclear charge radii, binding energies of doubly magic nuclei.
The forces SLy5+T and T44 involve the tensor terms added with
(T44) and without (SLy5+T) refitting the parameters of the central
interaction. These five parametrizations describe correctly the
subshell order near the Fermi level of $^{90,92}$Zr, $^{92,94}$Mo.
To see this, the calculated neutron and proton single-particle
energies for the case of $^{92}$Zr and the experimental
data~\cite{M79} are plotted in Fig.~1. The calculations with SGII
reproduce well the experimental data and other choices of the
Skyrme forces do not improve the agreement.

We employ the isospin-invariant pairing force~(\ref{pair}), with
the value $\rho_{0}$=0.16~fm$^{-3}$ of the nuclear saturation
density corresponding to the SGII force. The strength $V_{0}$ is
fitted to get a reasonable description of the experimental neutron
and proton pairing energies of $^{90,92}$Zr, $^{92,94}$Mo,
\begin{equation}
\label{P} P_{N}=\frac{1}{2}\left( B\left( N,Z\right) +B\left(
N-2,Z\right) -2B\left( N-1,Z\right) \right)
\end{equation}
for neutrons, and similarly for protons. Thus, the strength
$V_{0}$ is taken equal to -270 MeVfm$^{3}$, -420 MeVfm$^{3}$, and
-870 MeVfm$^{3}$ for the cases of the volume, the mixed and the
surface-peaked interaction, respectively. In order to make the
choice of the pairing interaction, it is very useful to analyze
the three lowest two-quasiparticle states in $^{92}$Zr. Since the
$2d_{5/2}$ neutron subshell is partially filled and the $2p_{1/2}$
proton subshell is filled, one can expect that the first level is
the $\{2d_{5/2},2d_{5/2}\}$ neutron state while the second level
is the $\{1g_{9/2},1g_{9/2}\}$ proton one, see for
example~\cite{LS04,LS06,pbl08}. As it is seen from  Fig.~2 the
different choices  have a strong impact on the unperturbed
two-quasiparticle excitations. Only the volume pairing interaction
gives the $\{1g_{9/2},1g_{9/2}\}$ proton state as the second
level. Thus, hereafter we use the Skyrme interaction SGII in the
particle-hole channel together with the volume zero-range force
acting in the particle-particle channel.

The Landau parameters $F^{ph}_0$, $F^{ph'}_0$ expressed in terms
of the Skyrme force parameters~\cite{sg81} depend on $k_F$. As it
is pointed out in our previous works~\cite{gsv98,ssvg02} one needs
to adopt some effective value for $k_F$ to give an accurate
representation of the original p-h Skyrme interaction. For the
present calculations we use the value $k_F=1.45$fm$^{-1}$ adjusted
so as to have the spurious center-of-mass state at zero energy.

It is worth to mention the effect of the residual p-p interaction
within the QRPA. Fig.~3 shows the effect on the $2^+_{1,2}$
energies and the $B(E2)$ values of $^{92}$Zr. Including the
quadrupole p-p interaction results in a decrease of the
$2^+_{1,2}$ energies (about 0.1~MeV) and the $B(E2)$-values do not
change practically.

Finally, we discuss the extension of the space for one- and
two-phonon configurations. To construct the wave
functions~(\ref{wf2ph}) of the low-lying  $2^+$ states up to 4~MeV
we use only the $2^+$ phonons, and all one- and two-phonon
configurations with energies up to 8~MeV are included. We have
checked that the inclusion of the high-energy configurations plays
a minor role in our calculations.
%
%
\section{Results}
\subsection{QRPA analysis}
\begin{table*} \caption{Energies, transition probabilities, and structures of the QRPA quadrupole
states in $^{90,92}$Zr and $^{92,94}$Mo. Phonon amplitude
contributions greater than 10\% are given.}
\begin{ruledtabular}
\begin{tabular}{ccccccccc}
 &State & Energy & $B(M1;2_{i}^+\rightarrow 2_1^+)$&$B(E2;0_{gs}^+\rightarrow 2_i^+)$&$\{n_1l_1j_1,n_2l_2j_2\}_{\tau}$& X & Y& $\%$        \\
 &      & (MeV)  &    ($\mu_N^2$)                 & (e$^2$fm$^4$)                   &                                &   &  &    \\
\noalign{\smallskip}\hline\noalign{\smallskip} $^{90}$Zr
 &$2_1^+$& 2.8 &     &630  &  $\{2d_{5/2},1g_{9/2}\}_\nu$   &-0.37 &-0.11 & 13           \\
 &       &     &     &     &  $\{1g_{9/2},1g_{9/2}\}_\pi$   & 1.03 & 0.06 & 53           \\
 &       &     &     &     &  $\{2p_{1/2},2p_{3/2}\}_\pi$   &-0.52 &-0.03 & 26           \\
\noalign{\smallskip}
 &$2_2^+$& 3.4 &0.00 &10   &  $\{2p_{1/2},2p_{3/2}\}_\pi$   & 0.79 & 0.00 & 63           \\
 &       &     &     &     &  $\{1g_{9/2},1g_{9/2}\}_\pi$   & 0.85 & 0.00 & 36           \\
\noalign{\smallskip}\hline\noalign{\smallskip} $^{92}$Zr
 &$2_1^+$& 1.7 &     &410  &  $\{2d_{5/2},2d_{5/2}\}_\nu$   & 1.26 & 0.12 & 79           \\
 \noalign{\smallskip}
 &$2_2^+$& 2.9 &0.53 &310  &  $\{2d_{5/2},2d_{5/2}\}_\nu$   &-0.63 & 0.09 & 20           \\
 &       &     &     &     &  $\{3s_{1/2},2d_{5/2}\}_\nu$   &-0.45 &-0.04 & 20           \\
 &       &     &     &     &  $\{1g_{9/2},1g_{9/2}\}_\pi$   & 0.85 & 0.04 & 36           \\
 &       &     &     &     &  $\{2p_{1/2},2p_{3/2}\}_\pi$   &-0.36 &-0.02 & 13           \\
\noalign{\smallskip}\hline\noalign{\smallskip} $^{92}$Mo
 &$2_1^+$& 1.9 &     &1170 &  $\{2d_{5/2},1g_{9/2}\}_\nu$   &-0.35 &-0.16 & 10           \\
 &       &     &     &     &  $\{1g_{9/2},1g_{9/2}\}_\pi$   & 1.32 & 0.19 & 86           \\
\noalign{\smallskip}
 &$2_2^+$& 4.4 &0.25 &230  &  $\{2d_{5/2},1g_{9/2}\}_\nu$   &-0.65 &-0.07 & 42           \\
 &       &     &     &     &  $\{2p_{1/2},2p_{3/2}\}_\pi$   &-0.63 &-0.02 & 40           \\
 &       &     &     &     &  $\{1g_{9/2},1g_{9/2}\}_\pi$   &-0.49 & 0.10 & 11           \\
\noalign{\smallskip}\hline\noalign{\smallskip} $^{94}$Mo
 &$2_1^+$& 1.2 &     &1730 &  $\{2d_{5/2},2d_{5/2}\}_\nu$   & 0.92 & 0.27 & 39           \\
 &       &     &     &     &  $\{1g_{9/2},1g_{9/2}\}_\pi$   & 0.99 & 0.37 & 42           \\
\noalign{\smallskip}
 &$2_2^+$& 2.4 &1.23 &160  &  $\{2d_{5/2},2d_{5/2}\}_\nu$   &-1.08 & 0.07 & 58           \\
 &       &     &     &     &  $\{1g_{9/2},1g_{9/2}\}_\pi$   & 0.89 & 0.02 & 40           \\
\end{tabular}
\end{ruledtabular}
\end{table*}
\begin{table*}
\caption{Energies, transition probabilities and dominant
components of phonon structures of the low-lying quadrupole states
in $^{90,92}$Zr and $^{92,94}$Mo. Experimental data are taken from
Refs.~\cite{P99,F03,F05,B00,G03}.}
\begin{ruledtabular}
\begin{tabular}{ccccccccc}
 &$\lambda_i^{\pi}=2_i^+$&\multicolumn{2}{c}{Energy}&Structure&\multicolumn{2}{c}{$B(E2;0_{gs}^+\rightarrow 2_i^+)$}&\multicolumn{2}{c}{$B(M1;2_{i}^+\rightarrow 2_1^+)$}\\
 &                       &\multicolumn{2}{c}{(MeV)} &         &\multicolumn{2}{c}{(e$^2$fm$^4$)}                  &\multicolumn{2}{c}{($\mu_N^2$)}\\
 &                       & Expt.  &  Theory         &         &  Expt.   &   Theory                               & Expt.          & Theory    \\
\noalign{\smallskip}\hline\noalign{\smallskip}$^{90}$Zr
 &$2_1^+$ &2.186&2.6& 93$\%[2_1^+]_{QRPA}$                                & 643$\pm$22      & 600   &                &               \\
\noalign{\smallskip}
 &$2_2^+$ &3.308&3.2& 95$\%[2_2^+]_{QRPA}$                                &  53$\pm$14      &   1   &0.088$\pm$0.025 & 0.00  \\
\noalign{\smallskip}\hline\noalign{\smallskip}$^{92}$Zr
 &$2_1^+$ &0.934&1.6& 96$\%[2_1^+]_{QRPA}$                                & 790$\pm$62      & 420   &                &               \\
\noalign{\smallskip}
 &$2_2^+$ &1.847&2.7& 87$\%[2_2^+]_{QRPA}$                                & 419$\pm$49      & 230   &  0.37$\pm$0.04 & 0.41   \\
\noalign{\smallskip}
 &$2_3^+$ &2.067&2.6&45$\%[2_4^+]_{QRPA}$+37$\%[2_1^+\otimes2_1^+]_{QRPA}$& $<0.62$         &  50   &  $<$0.024      & 0.17   \\
\noalign{\smallskip}\hline\noalign{\smallskip}$^{92}$Mo
 &$2_1^+$ &1.509&1.9& 99$\%[2_1^+]_{QRPA}$                                &1036$\pm$62      & 1160  &                &               \\
\noalign{\smallskip}
 &$2_2^+$ &3.091&3.8& 91$\%[2_1^+\otimes2_1^+]_{QRPA}$                    & 254$\pm$20      &   50  &0.043$\pm$0.007 & 0.03  \\
\noalign{\smallskip}\hline\noalign{\smallskip}$^{94}$Mo
 &$2_1^+$ &0.871&0.5& 73$\%[2_1^+]_{QRPA}$                                & 2031$\pm$25     & 1280  &                &               \\
\noalign{\smallskip}
 &$2_2^+$ &1.864&1.8&53$\%[2_1^+\otimes2_1^+]_{QRPA}$+21$\%[2_3^+]_{QRPA}$& 32$\pm$7        & 170   &  0.06$\pm$0.02 & 0.07   \\
\noalign{\smallskip}
 &$2_3^+$ &2.067&2.3& 87$\%[2_2^+]_{QRPA}$                                & 279$\pm$25      & 310   &  0.56$\pm$0.05 & 0.68  \\
\end{tabular}
\end{ruledtabular}
\end{table*}
First, properties of the low-lying quadrupole states are studied
within the one-phonon approximation. Results of our calculations
for the $2^+_{1,2}$ states in $^{90,92}$Zr, $^{92,94}$Mo: the
energies, the $B(E2)$, $B(M1)$ values, and the contributions
greater than 10\% to the normalization
conditions~(\ref{phononnorm}) are given in Table I. Note that the
$B(M1)$ values have been calculated with the $g$-factors of free
protons and neutrons.

One can see that we find a satisfactory description of the
isotopic dependence of the $2^+_1$ energies near closed shells.
The closure of the neutron subshell $1g_{9/2}$ in $^{90}$Zr,
$^{92}$Mo leads to the vanishing of the neutron pairing and as a
result energies of the first two-quasiparticle poles,
$\{1g_{9/2},1g_{9/2}\}_\pi$ in $^{90}$Zr, $^{92}$Mo are larger
than energies of the first poles, $\{2d_{5/2},2d_{5/2}\}_\nu$ in
$^{92}$Zr, $^{94}$Mo. This yields that the $2_1^{+}$ state has
collective (noncollective) structure with the domination of the
proton configuration $\{1g_{9/2},1g_{9/2}\}$ for the case of
$^{90}$Zr ($^{92}$Mo). On the other hand, in $^{92}$Zr the leading
neutron configuration $\{2d_{5/2},2d_{5/2}\}$ gives a contribution
of 79\% that is almost twice larger than in $^{94}$Mo. The
structure peculiarities are reflected in the
$B(E2;0_{gs}^+\rightarrow 2_1^+)$ values, as is shown in Table I.
The $2^+_1$ states in $^{92}$Zr, $^{94}$Mo have an isoscalar
character since the dominant neutron and proton phonon amplitudes
$X,Y$ of the $2^+_1$ state are in phase.

There are the second $2^+$ states in $^{92}$Zr, $^{94}$Mo below
3~MeV within the one-phonon approximation. The dominant neutron
and proton amplitudes of the fairly collective $2^+_2$ state are
out of phase. As a consequence, the isovector character of the
$2^+_2$ states are reflected in the remarkable values of
$B(M1;2^+_2\rightarrow 2^+_1)$, as given in Table I. The first
calculations of the isovector behavior of $2^+_2$ QRPA excitations
in $^{92}$Zr, $^{94}$Mo based on the analysis of the phonon
amplitudes within the QPM have been done in \cite{LS00,LS04,LS06}.

We turn now to the structures of the $2^+_2$ states of $^{90}$Zr,
$^{92}$Mo. The proton configurations exhaust about 99\% and 56\%
of the wave function normalization in $^{90}$Zr and $^{92}$Mo,
respectively. It means that the second pole,
$\{2p_{1/2},2p_{3/2}\}_\pi$ in $^{92}$Mo is closer to the neutron
poles than that in $^{90}$Zr. As expected, the negligible size of
the $B(M1;2^+_2\rightarrow 2^+_1)$ value of $^{90}$Zr is obtained.
The $2^+_2$ state in $^{92}$Mo has the main neutron and proton
phonon amplitudes in phase (Table I) and this results in the
comparable $B(M1)$ value of the $M1$ transitions between the
noncollective $2^+_1$ state and the isoscalar $2^+_2$ state.

This analysis within the one-phonon approximation can help to
identify the MS state, but it is only a rough estimate. Some
overestimate of the experimental energies (Table II) indicates
that there is room for two-phonon effects.
%
%
\subsection{Effects of the phonon-phonon coupling}
Let us now discuss the extension of the space to one- and
two-phonon configurations. The calculated $2^+$ state energies,
the largest contributions of the wave function
normalization~(\ref{norma2ph}), the $B(E2)$, $B(M1)$ values and
the experimental data~\cite{P99,F03,F05,B00,G03} are shown in
Table II.  It is worth pointing out that we get the wrong energy
order of the second and third $2^+$ states in the case of
$^{92}$Zr. On the other hand, the energy difference equal to
0.1~MeV is close to the expected accuracy of our calculations.

One can see that the inclusion of the two-phonon terms results in
a decrease of the $2^+_1$ energies and in a reduction of the
$B(E2;0_{gs}^+\rightarrow 2_1^+)$ values, except for $^{92}$Zr.
Our calculations reproduce well a general behavior for energies
and transition probabilities. There is some underestimation of the
$B(E2)$ values of the N=52 isotones in comparison with the
experimental data, this probably points to a particular problem
due to the effective interaction rather than to a deficiency of
our variational space. In all four nuclei, the crucial
contribution in the wave function structure of the first $2^+$
state comes from the $[2^+_1]_{QRPA}$ configuration, but the
two-phonon contributions are appreciable. This means that the
structures of the first $2^+$ states do not change practically due
to the effects of the phonon-phonon coupling. As a result, we get
the isoscalar collective structure of the $2^+_1$ state in
$^{94}$Mo and the neutron dominated $2^+_1$ excitation of
$^{92}$Zr which indicates the F-spin breaking.

The second $2^{+}$ state contains a dominant two-phonon
configuration $[2_1^+\otimes2_1^+]_{QRPA}$ in $^{92,94}$Mo and
such contribution leads to the small $B(E2)$ values. Moreover, the
calculated $B(M1)$ value is sensitive to the phonon composition:
the two-phonon configuration composed of the isoscalar collective
phonons in the case $^{94}$Mo and the noncollective phonons of
$^{92}$Mo. In $^{90,92}$Zr, the wave function of the $2^+_2$ state
is dominated by the $[2^+_2]_{QRPA}$ configuration and we can
follow the QRPA estimate discussed in Sec.~IVA. For the case of
$^{92}$Zr we obtain the isovector collective $2^+_2$ state and the
dominant one-phonon structure of the $M1$ transition between the
first and second $2^+$ states in $^{92}$Zr. The results are close
to those that were previously calculated in the shell
model~\cite{W02,F05,H07} and in the QPM~\cite{LS04,LS06,W11}.

Finally, we examine the occurrence at low energy (below 3 MeV) of
the third $2^+$ states. One can see that the collective state in
$^{94}$Mo is dominated by the isovector one-phonon
$[2_2^+]_{QRPA}$ structure. The calculated values of the $2^+_3$
energy and the transition probabilities are in reasonable
agreement with the experimental data. In other words, we reproduce
the IBM-2 level scheme and the calculated $B(M1;2^+_3\rightarrow
2^+_1)$ value supports the MS assignments observed experimentally
and theoretically for the first time in Ref.~\cite{P99}. It is
noteworthy that this conclusion for $^{94}$Mo remains valid for
the SLy5+T parameter set~\cite{savpg11}. For the $2^+_3$ state of
$^{92}$Zr, one of the main components of the wave function is the
$[2_1^+\otimes2_1^+]_{QRPA}$ configuration. As can be seen from
Table II, the two-phonon contribution is reflected in the small
values of the transition probabilities, but there is some
overestimation in comparison with the experimental data. One can
expect an improvement if the variational space is enlarged by the
phonon composition with hexadecapole multipolarity and the
three-phonon configurations are taken into account. Such
calculations are now in progress.

Thus the microscopic approach~\cite{gsv98,svg08,svg04} describes
the properties of the low-lying states in $^{92}$Zr, $^{94}$Mo
less accurately than more phenomenological
ones~\cite{B07,LS00,LS04,LS06}, but the results are still in a
reasonable agreement with the experimental
data~\cite{P99,F03,F05,B00,G03}.
%
%
\section{Conclusions}
Starting from the Skyrme mean-field calculations, we have studied
the effects of the phonon-phonon coupling on the properties of the
low-energy spectrum of $2^+$ excitations and, in particular, on
the $M1$ transitions between the excited states of nuclei in the
mass range $A\approx90$. The finite-rank separable approach for
the QRPA enables one to perform the calculations in very large
configurational spaces.

The parametrization SGII of the Skyrme interaction is used for all
calculations in connection with the volume zero-range pairing
interaction. Using the same set of parameters we have studied the
behaviour of the energies, the $B(E2;0_{gs}^+\rightarrow 2_i^+)$
and $B(M1;2_{i}^+\rightarrow 2_1^+)$ values of the lowest $2^+$
states in $^{90,92}$Zr, $^{92,94}$Mo. Among our initial motivation
was the search for MS states in $^{92}$Zr, $^{94}$Mo in comparison
to the $N=50$ isotones $^{90}$Zr, $^{92}$Mo with closed neutron
shell. Our results indicate indeed the occurrence of MS states in
our calculation for the nuclei $^{92}$Zr and $^{94}$Mo that were
successfully predicted within the IBM-2 before. Our results from
the Skyrme interaction are in reasonable agreement with
experimental data. We stress that they represent the first
successful comparison between experimental $M1$ transition values
and those calculated with the Skyrme interaction. The coupling
between one- and two-phonon terms in the wave functions of excited
states is essential. The QRPA results tend to overestimate the
$2^+_1$ energies and the inclusion of the two-phonon
configurations results in a decrease of the energies. There is a
clear influence on the structure of the $2^+_{2,3}$ states. The
structure of the low-lying $2^{+}$ states calculated in our
approach are close to those that were calculated within the QPM
before. We conclude that the present approach may provide a
valuable globally applicable microscopic analysis of the
properties of the lowest quadrupole excitations.

Our model would probably be improved by enlarging the variational
space for the $2^{+}$ states with the inclusion of the two-phonon
configurations constructed from phonons with hexadecapole
multipolarity and taking into account the three-phonon
configurations. The computational developments that would allow us
to conclude on this point are still underway.
%
%
\section*{Acknowledgments}
We are grateful to R.V. Jolos, Nguyen Van Giai, V.Yu. Ponomarev,
Ch. Stoyanov and V. V. Voronov for useful discussions. A.P.S. and
N.N.A. thank the hospitality of Institut f\"{u}r Kernphysik,
Technische Universit\"{a}t Darmstadt where a part of this work was
done. This work was partly supported by the Heisenberg-Landau
program, by the DFG under grant No.~SFB634, and by the RFBR grant
No.~110291054.
%
%

%
%
\end{document}